\documentclass[showpacs,preprintnumbers,amsmath,amssymb,superscriptaddress]{revtex4}
\usepackage{epsfig}
\usepackage{psfrag}
\usepackage{amsfonts}
\usepackage{graphicx}
\usepackage{dcolumn}
\usepackage{bm}
\usepackage{xcolor}
\begin{document}


\title{Magnetic field effect on neutral color superconductivity as a mechanism to boost the inner magnetic field of magnetars }


\author{Bo  Feng}
\affiliation{
School of Physics, Huazhong University of Science and Technology, Wuhan 430074, China
}

\date{\today}

\begin{abstract}
In this paper, we review the impact of a strong magnetic field on the neutral two-flavor color superconducting (2SC) phase of dense quark matter. Using a Nambu-Jona-Lasinio (NJL) model approach in the mean-field approximation, we first solve the gap equations for the diquark condensate $\Delta$ and the electric chemical potential $\mu_e$ under a uniform external magnetic field, and discuss the coupled evolution of the chiral and diquark order parameters. We then evaluate the field-dependent Meissner masses of the rotationally charged gluons for a wide range of densities and moderate diquark coupling strengths. Combining the "zero-mode problem" mechanism with the neutrality condition, we show that the gluon masses become tachyonic even at moderate magnetic-field values, signaling a chromomagnetic instability in density regions where the system was stable at zero field. This instability can be resolved through the restructuring of the ground state with the formation of a gluon vortex condensate. A remarkable consequence of the vortex configuration is the paramagnetic amplification of the in-medium magnetic field: seed fields of the order of $10^{15}$ G can be enhanced by up to two orders of magnitude, reaching $\sim 10^{17}$ G, precisely the range predicted for magnetar cores. We compare this microscopic mechanism with the conventional dynamo explanation and discuss how it offers a plausible explanation for the extremely high magnetic fields expected to be realized in magnetar interiors.
\end{abstract}

\keywords{quark matter; color superconductivity; neutron stars; magnetic field; finite-temperature field theory}
\maketitle


\section{Introduction}

The properties of strongly interacting matter at densities of and beyond nuclear saturation density and temperatures ranging from zero up to hundreds MeV are subject to intense investigations \cite{From_hadrons_to_quarks}, because these conditions are relevant for both baryon rich heavy ion collisions \cite{HIC, HIC-2} and the interior of neutron stars (NS) \cite{Sedrakian,Weber, Quark-stars,Fukushima-QCD-astrophysics}.  In this dense regime, neither perturbative QCD nor lattice calculations are reliable, making a full understanding  of the  physical properties of matter under these conditions particularly challenging.  Nonetheless, various QCD-inspired model studies suggest that the quark matter at high density may exhibit a very rich phase structure. With the detection of gravitational waves from a binary NS merger \cite{GW170817_2, GW170817}, a new observational window has opened for studying the properties of dense quark matter. Hopefully, combining observational data with theoretical studies may shed new light on the physics under such extreme conditions.

A dense Fermi system at sufficiently low temperature is unstable against Cooper pairing for an arbitrarily weak attractive interaction \cite{BCS}. In QCD, such pairing can be triggered by attractive interactions from either perturbative one-gluon exchange at asymptotically high baryon density or instanton effects at moderate density \cite{Barrios,Bailin-Love,Iwasaki-Iwado,Nucloen-droplets-and-CS,Schafer-Shuryak,CFL,Berges-Rajagopal,Condensed_qcd,CS-reveiw}. Once quarks are deconfined, a color superconducting (CS) phase is expected to form in the cold and dense environment, characterized by Cooper pairs of quarks with opposite momentum and spin. When the $u$, $d$, and $s$ quark masses are negligible compared to the baryon chemical potential, all three flavors pair, resulting in the spin-zero color-flavor-locked (CFL) phase \cite{CFL}, which is generally accepted as the ground state at asymptotically high density and low temperature. At moderate densities, if the density is high enough to deconfine quarks but low enough to decouple the strange quark, only the two light flavors pair, and the ground state is the two-flavor superconducting (2SC) phase.


On the other hand, an important feature of NS's is their strong magnetization. To date, over 4,000 pulsars have been discovered \cite{FAST}, and observational surveys have revealed a clear distinction among different populations. Normal pulsars have surface magnetic fields concentrated around $10^{12}$~G \cite{NS-B-1}, while millisecond pulsars (MSPs) possess much weaker fields, typically in the range of $10^8$--$10^9$~G \cite{MillisecondPulsasr, MillisecondPulsasr-2}, with most distributed below $10^{10.5}$~G. In contrast, a class of neutron stars known as magnetars \cite{NS-B-2, Thompson}, inferred from spectroscopic and spin-down studies of soft gamma-ray repeaters (SGRs) and anomalous X-ray pulsars (AXPs), possess significantly stronger surface fields, estimated to lie between $10^{14}$ and $10^{15}$~G \cite{Turolla-Zane, Kaspi}. Due to magnetic flux conservation in highly conductive stellar matter, the core field of magnetars could be substantially larger, with theoretical estimates suggesting $10^{17}\sim 10^{20}$ G \cite{magnetizedfermions,Ferrer-Keith-Portillo,Uryu, Uryu_2, Sobyanin,Rea}. This has motivated numerous studies of the equation of state (EOS) of magnetized NS \cite{insignificance,BF-spin-one, E-V-Review,Theor-estimations,Chatterjee,Ferrer-Hackebill}. The origin of these huge fields, however, remains an open question.

In the 2SC phase, the original electromagnetic $U(1)_{em}$ symmetry is broken, but a remnant $\widetilde{U}(1)$ symmetry persists. The corresponding gauge field is a mixture of the photon field and the eighth gluon field, $\widetilde{A}_\mu = \cos\theta A_\mu - \sin\theta G^8_\mu$, with $\tan\theta = e/(\sqrt{3}g)$, where $e$ and $g$ are the electromagnetic and strong coupling constants respectively \cite{Rotated Fields, Rotated_2}. Because this combination remains massless, the rotated magnetic field can penetrate the stellar medium without being screened by the Meissner effect. This rotated electromagnetic field assigns charges to gluons. Since gluons are spin-1 fields, a sufficiently strong rotated magnetic field can trigger an instability through the so-called ``zero-mode problem''. The ground state is then restructured by charged gluon vortices, which simultaneously boost the magnetic field \cite{MCFL-Vortices,Ferrer-Vortices1, MCFL-Vortices-2,Ferrer-Abrikosov-Vortex}. Another factor affecting the stellar medium is the condition of neutrality. The requirement that quark matter inside NSs be electrically and color neutral and in \(\beta\)-equilibrium leads to a mismatch \(\delta\mu_e = \mu_e/2\) between the Fermi surfaces of pairing quarks \cite{Rischke-Debye-Meissner,Rischke-Shovkovy,Huang-Igor,Igor, Schmitt-Wang-Rischke}. For \(0<\Delta/\delta\mu_e<1\), gapless modes appear, giving rise to the gapless 2SC (g2SC) phase. In this phase, the squared Meissner mass of charged gluons can become negative, signaling a chromomagnetic instability \cite{Huang-Igor-1, Huang-Igor-2}. This instability persists even in the gapped 2SC phase for \(1<\Delta/\delta\mu_e<\sqrt{2}\). Possible resolutions include inhomogeneous phases with momentum-dependent gaps \cite{Crystalline-CFL,Reddy,Ren-LOFF,Ren-LOFF-1,Ren-LOFF-2,Miransky-LOFF, Fukushima,Hashimoto}, based on the ideas of Larkin and Ovchinnikov \cite{Larkin} and Fulde and Ferrell \cite{Ferrell}, although a consensus has not yet been reached on how to remove this instability.

In the present review, we will focus on the possibility of an alternative scenario involving gluon-vortex condensates, which will serve as essential factors for eliminating the chromomagnetic instability in neutral 2SC under a magnetic field. This scenario is motivated by the solution to the "zero-mode problem" mentioned above-a mechanism that has been explored in various contexts such as Yang-Mills theory \cite{Yang-Mills,Yang-Mills_2}, electroweak theory \cite{W-EW1, W-EW2,W-EW-2,W-EW-3}, color superconductivity \cite{MCFL-Vortices,MCFL-Vortices-2,Ferrer-Vortices1} and even in string theory \cite{String Theory, String-2}.  While previous studies have examined the neutral 2SC phase in the presence of a magnetic field, they were primarily limited  to analyzing the behavior of the quark gap $\Delta$ and constituent quark mass $m$ as function of density and magnetic field \cite{Neutral-2SC-Magnetic-Field,Jaikumar-1,Jaikumar-2,Jaikumar-3,Scoccola,Mishra}.  To date, the impact of an external magnetic field on the gluon Meissner mass of such an electrically neutral system has only been addressed in \cite{BF-Ef} showing that in the presence of a magnetic field the unstable neutral 2SC phase can be stabilized by restructuring its ground state involving the formation of a gluon vortex condensate.  There, it was also shown that once this gluon vortex condensate develops the medium will exhibit paramagnetic behavior, significantly amplifying the in-medium magnetic field. This amplification mechanism offers a plausible explanation for the extremely strong inner magnetic field in magnetars. 


In this paper, we review several previous results related to the role of a magnetic field in the chromomagnetic instability of neutral 2SC matter. The paper is organized as follows. In Section II, we present the NJL formulation of the neutral 2SC phase in a uniform magnetic field. In Sec. II.A, we solve the gap equations for the diquark condensate $\Delta$ and the electric chemical potential $\mu_e$. In Sec. II.B, we discuss the coupled evolution of the chiral and diquark order parameters. In Sec. II.C, we compute the Meissner masses of the charged gluons, demonstrating that for moderate couplings they become tachyonic even at moderately high magnetic fields. In Section III, we discuss how this instability can be resolved. In Sec. III.A, we review the zero-mode problem and the formation of a gluon vortex condensate. In Sec. III.B, we introduce the gluonic phase as a more general Ginzburg-Landau framework with vector condensates of gluons. In Section IV, we show that the gluon vortex condensate induces a paramagnetic response that significantly amplifies the in-medium magnetic field, and discuss how this amplification mechanism can serve as a possible explanation for the extreme magnetic fields observed in magnetars. Finally, Section V presents our concluding remarks.

\section{Neutral 2SC in a magnetic field}

\subsection{General formalism}
The study of neutral 2SC quark matter in the presence of an external magnetic field was first investigated in \cite{Neutral-2SC-Magnetic-Field,Jaikumar-1,Jaikumar-2,Jaikumar-3,Scoccola,Mishra} with emphasis on the behavior of the gap and the constituent mass. Here we will review in detail the formalism and numerical results ignoring the mass, since in the density regime relavant for NS the constituent mass already vanishes. We start with a two flavor NJL model of massless quarks in the presence of an external constant and uniform rotated magnetic field. The Lagrangian reads
\begin{align}
{\cal L}=\bar\psi\left[i\gamma^\mu(\partial_\mu+i\tilde{e}\tilde{Q}\tilde{A}_\mu)+\mu\gamma^0\right]\psi+G_S\left[(\bar\psi \psi)^2+(\bar\psi i\gamma^5{\bf \tau}\psi)^2\right]+G_D\left(\bar\psi i\gamma^5T^2\tau^2\psi^C \right)\left(\bar\psi^Ci\gamma^5T^2\tau^2\psi\right).\label{Lagrangian}
\end{align}
Here, $\psi^T=(u,d)$ denotes the quark doublet, while $\psi^C=C\bar\psi^T$ is its charge conjugate, with $C=i\gamma^2\gamma^0$ being the charge conjugation operator. The four-fermion interaction contains scalar and diquark channels with coupling constants $G_S$ and $G_D$, respectively. In color space, $T^2$ represents the antisymmetric second Gell-Mann matrix, and $\tau^2$ is the antisymmetric Pauli matrix in flavor space. We adopt the Landau gauge for the external magnetic field, where the electromagnetic potential reads $\tilde{A}^\mu=(0,0,\tilde{B}x,0)$. This corresponds to a constant and uniform magnetic field $\tilde{B}$ along the positive $z$ direction; the wavy-hat notation is introduced to distinguish these quantities from the conventional magnetic field. The rotated charge matrix $\tilde{Q}$ is defined as
$\tilde{Q}=Q_f\times 1_c-1_f\times \frac{T_c^8}{2\sqrt{3}}$. 
We adopt units in which the rotated charge is normalized as $\tilde{e}=\frac{\sqrt{3}ge}{\sqrt{3g^2+e^2}}>0$. The original electromagnetic charge matrix for quarks is $Q_f={\rm diag}(2/3, -1/3)$, while $T_c^8={\rm diag}(1/\sqrt{3}, 1/\sqrt{3},-2/\sqrt{3})$ denotes the eighth Gell-Mann matrix. The corresponding $\tilde{Q}$ charges for quarks and gluons are listed explicitly in Tables I and II.
 \begin{table}[htbp]
	\centering
	\caption{$\tilde{Q}$ charge for different quarks}
	\begin{tabular}{cccccc}
		\hline
		$u_r$\ \ \ \ & $u_g$\ \ \ \ & $u_b$\ \ \ \ & $d_r$\ \ \ \ & $d_g$\ \ \ \ & $d_b$  \\
		\hline
		$\frac{1}{2}$\ \ \ \ & $\frac{1}{2}$\ \ \ \ & 1\ \ \ \ & $-\frac{1}{2}$\ \ \ \ & $-\frac{1}{2}$\ \ \ \ & 0  \\
		\hline
	\end{tabular}
\end{table}
\begin{table}[htbp]
	\centering
	\caption{$\tilde{Q}$ charge for different gluons}
	\begin{tabular}{cccccccc}
		\hline
		$G_\mu^1$\ \ \ \ & $G_\mu^2$\ \ \ \ & $G_\mu^3$\ \ \ \ & $G_\mu^+$\ \ \ \ & $G_\mu^-$\ \ \ \ & $H_\mu^+$\ \ \ \ & $H_\mu^-$ \ \ \ \ & ${\tilde G}_\mu^8$\\
		\hline
		$0$\ \ \ \ & $0$\ \ \ \ & $0$\ \ \ \ & $\frac{1}{2}$\ \ \ \ & $-\frac{1}{2}$\ \ \ \ & $\frac{1}{2}$\ \ \ \ & $-\frac{1}{2}$\ \ \ \ &0\\
		\hline
	\end{tabular}\label{chargedgluons}
\end{table}
We note that the interaction term in the Lagrangian allows only the diquark pairs $u_rd_g$ and $u_gd_r$, which carry zero $\tilde{Q}$ charge and are therefore consistent with the unbroken ${\tilde U}(1)$ symmetry. The charged gluon fields are linear combinations of $G^{4,5}_\mu$ and $G^{6,7}_\mu$，
\begin{equation}
G_\mu^\pm=\frac{1}{\sqrt{2}}(G_\mu^4\mp iG_\mu^5), \ \ \ \ \ \ H_\mu^\pm=\frac{1}{\sqrt{2}}(G_\mu^6\mp iG_\mu^7).
\end{equation}

The quark chemical potentials that enforce electric and color neutrality, together with $\beta$-equilibrium, can be written in color-flavor space as
\begin{equation}
\mu_{ij,\alpha\beta}=\left[\mu\delta_{ij}-\mu_e(Q_f)_{ij}\right]\delta_{\alpha\beta}+\frac{2}{\sqrt{3}}\mu_8\delta_{ij}(T^8)_{\alpha\beta}.
\end{equation}
For individual quark flavors and colors, the components read explicitly as
\begin{equation}
\mu_{u_r}=\mu_{u_g}=\bar\mu-\delta\mu_e,
\end{equation}
\begin{equation}
\mu_{d_r}=\mu_{d_g}=\bar\mu+\delta\mu_e,
\end{equation}
\begin{equation}
\mu_{u_b}=\bar\mu-\delta\mu_e-\delta\mu_8,
\end{equation}
\begin{equation}
\mu_{d_b}=\bar\mu+\delta\mu_e-\delta\mu_8.
\end{equation}
Here we have introduced the abbreviations $\bar{\mu}=\mu-\frac{1}{6}\mu_{e}+\frac{1}{3}\mu_{8}$, $\delta\mu_e=\frac{1}{2}\mu_{e}$, and $\delta\mu_8=\mu_{8}$. The parameters $\mu_e$ and $\mu_8$ are not fixed externally; they are determined dynamically by the requirements of electric and color neutrality. As shown in \cite{Igor, Jaikumar-2}, the color chemical potential $\mu_8$ is typically much smaller than $\mu_e$ and $\mu$, and has negligible impact on the results. We therefore set $\mu_8=0$ in what follows.

The partition function is given by
\begin{align}
{\cal Z}=\int{\cal D}\bar\Psi{\cal D}\Psi\exp({{\cal S}_{\rm eff}}),
\end{align}
where the Euclidean effective action ${\cal S}_{\rm eff}$ is
\begin{align} 
{\cal S}_{\rm eff} =\int_0^\beta d\tau\int d^3{\bf x}\left\{ \bar\psi\left[-\gamma^0\frac{\partial}{\partial \tau}+\gamma\cdot(i\nabla+\tilde{e}\tilde{Q}\tilde{\bf A})+\mu\gamma^0\right]\psi
+\frac{1}{2}\Delta\bar\psi i\gamma^5T^2\tau^2\bar\psi^C+\frac{1}{2}\Delta^*{\bar\psi}^Ci\gamma^5T^2\tau^2\psi-\frac{|\Delta|^2}{4G_D}\right\},
\end{align}
with $\Delta=2G_D\langle {\bar\psi}^Ci\gamma^5T^2\tau^2\psi \rangle$ defined in the mean-field approximation. We neglect the chiral condensate, since in the density range of interest for neutron stars ($\mu > 300$ MeV) it is expected to vanish. A detailed discussion of its interplay with the diquark condensate is given in Sec.II B.

For fields carrying ${\tilde Q}=\pm\frac{1}{2}$, it is convenient to introduce the Nambu-Gorkov spinors
\begin{align}
&\Psi_{({\tilde Q})}=\left(\begin{array}{c}\psi_{({\tilde Q})}\\ \psi^{C}_{(-{\tilde Q})}\end{array}\right),\quad\bar{\Psi}_{({\tilde Q})}=\left(\begin{array}{cc}\bar{\psi}_{({\tilde Q})}&\bar{\psi}^{C}_{(-{\tilde Q})}\end{array}\right),
\end{align}
while for  ${\tilde Q}=0, 1$
\begin{align}
&\Psi_{({\tilde Q})}=\psi_{({\tilde Q})},\quad\bar{\Psi}_{({\tilde Q})}=\bar{\psi}_{({\tilde Q})},
\end{align}
since those quarks do not participate in pairing. Here, we defined $\psi_{({\tilde Q})}=\Omega_{({\tilde Q})}\psi$ with charge projectors for the field representation $\psi^T=(u_r,u_g,u_b,d_r,d_g,d_b)$ given by
\begin{equation}
\Omega_{(0)}={\rm diag}(0,0,0,0,0,1),
\end{equation}
\begin{equation}
\Omega_{(\frac{1}{2})}={\rm diag}(1,1,0,0,0,0),
\end{equation}
\begin{equation}
\Omega_{(-\frac{1}{2})}={\rm diag}(0,0,0,1,1,0),
\end{equation}
\begin{equation}
\Omega_{(1)}={\rm diag}(0,0,1,0,0,0),
\end{equation}
in color and flavor space. Those projectors satisfy
\begin{equation}
\Omega_{\eta}\Omega_{\eta^\prime}=\delta_{\eta\eta^\prime}\Omega_{\eta}, \ \ \ \ \ \ \ \ \eta, \eta^\prime=0,\pm\frac{1}{2}, 1,
\end{equation}
and
\begin{equation}
\sum_\eta\Omega_{\eta}=1.
\end{equation}

The effective action can be written as
\begin{equation}
{\cal S}_{\rm eff}=\int_0^\beta d\tau\int d^3{\bf x}\left[\sum_{{\tilde Q}}{\bar\Psi}_{({\tilde Q})}(x)S_{({\tilde Q})}^{-1}(x)\Psi_{({\tilde Q})}(x)-\frac{\Delta^{2}}{4G_D}\right],    \label{effectiveaction2}     
\end{equation}
where the inverse propagators for charged fields with ${\tilde Q}=\pm\frac{1}{2}$ are given by
\begin{equation}
S_{({\tilde Q})}^{-1}(x)=\frac{1}{2}\left(\begin{array}{cc}
[G^+_{({\tilde Q})0}(x)]^{-1} & \Phi^-_{({\tilde Q})}\\
\Phi^+_{({\tilde Q})} & [G^-_{({\tilde Q})0}(x)]^{-1}\\
\end{array}\right),
\end{equation}
with
\begin{equation} 
[G^\pm_{\left(\tilde Q\right)0}(x)]^{-1}=-\gamma^0\frac{\partial}{\partial \tau}+\gamma\cdot\left( i\nabla\pm{\rm sgn}({\tilde Q})\frac{1}{2}|{\tilde e}{\tilde {\bf A}}|\right)\pm\gamma^{0}{\bar\mu}_{-{\rm sgn}({\tilde Q})},
\end{equation}
and
\begin{equation}
\Phi^+_{(\tilde Q)}={\rm sgn}({\tilde Q})\left(\begin{array}{cc}
0 & \Delta i\gamma^5\\
-\Delta i\gamma^5 & 0\\
\end{array}\right), \ \ \ \ \Phi^-_{({\tilde Q})}=\gamma^0\left[\Phi^+_{({\tilde Q})}\right]^\dagger\gamma^0.
\end{equation}
Whereas the inverse propagators for charged field with ${\tilde Q}=1$ and the neutral field are the conventional ones
\begin{equation}
S^{-1}_{(0)}(x)=-\gamma^0\frac{\partial}{\partial \tau}+\gamma\cdot i\nabla+\gamma^{0}{\bar\mu}_+,
\end{equation}
and
\begin{equation}
S^{-1}_{(1)}(x)=-\gamma^0\frac{\partial}{\partial \tau}+\gamma\cdot\left( i\nabla+|{\tilde e}{\tilde {\bf A}}|\right)+\gamma^{0}{\bar\mu}_-,
\end{equation}
with ${\bar\mu}_\pm\equiv {\bar\mu}\pm\delta\mu_e$.

The transformation to momentum space of the effective action (\ref{effectiveaction2}) can be performed by the Ritus's approach, which was widely used in the literatures when considering the dynamics of charged fermion fields as well as vector field in an external magnetic field (see \cite{Ritus,Ritus_approach,Ritus_approach-2,Ritus_approach-3,Ritus_approach-4,Ritus_approach-5,Ritus_approach-6,MCFL,MCFL-2,Ritus-Trans,Bo} for details). We therefore have
\begin{equation}
{\cal S}_{\rm eff}=\sum_{\tilde Q}\int\!\!\!\!\!\!\!\!\sum\frac{d^4p}{(2\pi)^4}{\bar\Psi}_{(\tilde Q)}(p){ S}^{-1}_{(\tilde Q)}(p)\Psi_{(\tilde Q)}(p)-\beta V\frac{\Delta^{2}}{4G_D},
\end{equation}
where
\begin{equation}
{S}^{-1}_{(\tilde Q)}(p)=\frac{1}{2}\left(\begin{array}{cc}
[G^+_{(\tilde Q)0}(p)]^{-1} & \Phi^-_{(\tilde Q)}\\
\Phi^+_{(\tilde Q)} & [G^-_{(\tilde Q)0}(p)]^{-1}\\
\end{array}\right)\Pi(l).
\end{equation}
for ${\tilde Q}=\pm\frac{1}{2}$. We have defined  $\Pi(l)=\Delta(\pm)\delta^{l0}+(1-\delta^{l0})$ with $\Delta(\pm)=(1\pm i\gamma^1\gamma^2)/2$ the spin projectors and $l=0,1,2,...$ the Landau levels. The diagonal elements are
\begin{equation}
[G^\pm_{(\tilde Q)0}(p)]^{-1}=\gamma_\mu{ p}^{\mu,\pm}_{(\tilde Q)}\pm\gamma^{0}{\bar\mu_{(-{\rm sgn} (\tilde Q))}},
\end{equation}
with ${ p}^{\mu,\pm}_{({\tilde Q})}=(i\omega_n, 0, \pm{\rm sgn}(\tilde Q)\sqrt{2|{\tilde e}{\tilde Q}{\tilde{\bf B}}|l},p^3)$.
While for ${\tilde Q}=1$
\begin{equation}
S^{-1}_{(1)}(p)=\gamma_\mu{ p}^{\mu}_{(1)}\pm\gamma^{0}{\bar\mu_-},
\end{equation}
and for ${\tilde Q}=0$
\begin{equation}
S^{-1}_{(0)}(p)=\gamma_\mu{ p}^\mu\pm\gamma^{0}{\bar\mu}_+.
\end{equation}

Integrating the NG spinors, we obtain the thermodynamical potential as
\begin{align}
\Omega=\frac{\Delta^{2}}{4G_D}-T\sum_{\tilde Q}\int\!\!\!\!\!\!\!\!\sum\frac{d^4 p}{(2\pi)^4}\frac{1}{2}{\rm Tr}\ln \left[S_{({\tilde Q})}^{-1}(p)\right]-\frac{1}{12\pi^2}\left(\mu_e^4+2\pi^2T^2\mu_e^2+\frac{7\pi^4}{15}T^4\right).
\end{align}
Note that we take into account the contribution from the electrons required to maintain the system neutral and $\beta$ equilibrated. 
Carrying out the sum in Mastubara frequencies,  we finally have
\begin{align}
\nonumber \Omega=&-\frac{1}{12\pi^2}\left(\mu_e^4+2\pi^2T^2\mu^2+\frac{7\pi^4}{15}T^4\right)+\frac{\Delta^2}{4G_D}-\int\frac{d^3{\bf p}}{(2\pi)^3}\sum_{\sigma=\pm}\left(E_{(0)}^\sigma+2T\ln\left[1+e^{-\frac{ E_{(0)}^\sigma}{T}}\right]\right)\\
&-\frac{|{\tilde e}{\tilde Q}{\tilde {B}}|}{(2\pi)^2}\int dp_3\sum_{l=0}^\infty(2-\delta_{l,0})\sum_{{\tilde Q}=1, \pm\frac{1}{2}}\sum_{\sigma, \xi=\pm}\left(E_{({\tilde Q})\xi}^\sigma+2T\ln\left[1+e^{-\frac{E_{(\tilde Q)\xi}^\sigma}{T}}\right]\right),
\end{align}
where $E_{\tilde Q}$ are the spectrum of the quasiparticles given with their corresponding degeneracies by
\begin{align}
E_{(0)}^\pm=&\pm\mu_{d_b}+|{\bf p}|\ \ \ \ \ \ \ \   \   (\times 2),\\
E_{(1)}^\pm=&\pm\mu_{u_b}+|{\bar{\bf p}}_{(1)}|\ \ \ \ \ \ (\times 2),\\
E_{(\frac{1}{2})\pm}^\pm=&\pm\delta\mu_e+\sqrt{(|{\bar{\bf p}}_{(\frac{1}{2})}|\pm\bar{\mu})^{2}+\Delta^{2}},\\
E_{(-\frac{1}{2})\pm}^\pm=&\pm\delta\mu_e+\sqrt{(|{\bar{\bf p}}_{(-\frac{1}{2})}|\pm\bar{\mu})^{2}+\Delta^{2}}, \label{eigenmodes}
\end{align}  
 with ${\bar{\bf p}}_{{(\tilde Q)}}=(0, {\rm sign}({\tilde Q})\sqrt{2|{\tilde e}{\tilde Q}{\tilde {B}}|}, p^3)$.  In this work, $\mu_e$ is treated as a dynamical parameter, determined by minimizing the thermodynamic potential together with the gap equation for $\Delta$,
 \begin{equation}\label{gap-eq}
 \frac{\partial\Omega}{\partial\Delta}=\frac{\partial\Omega}{\partial\mu_e}=0,
 \end{equation}
which we solve numerically. The NJL model parameters, i.e., the three-momentum cutoff $\Lambda$ and the scalar coupling $G_S$, are chosen as $\Lambda=653.3$ MeV and $G_S\Lambda^2=2.14$, fixed by fitting the experimental pion mass and decay constant \cite{Klevansky-Rep,Buballa-Rep}. The diquark coupling $G_D$, however, is not fixed by any phenomenologies. We simply adopt the convention $G_D=\eta G_S$ with $\eta=0.75, 0.85, 0.95$ and $1.15$.

\begin{figure}
	\includegraphics[height=10cm]{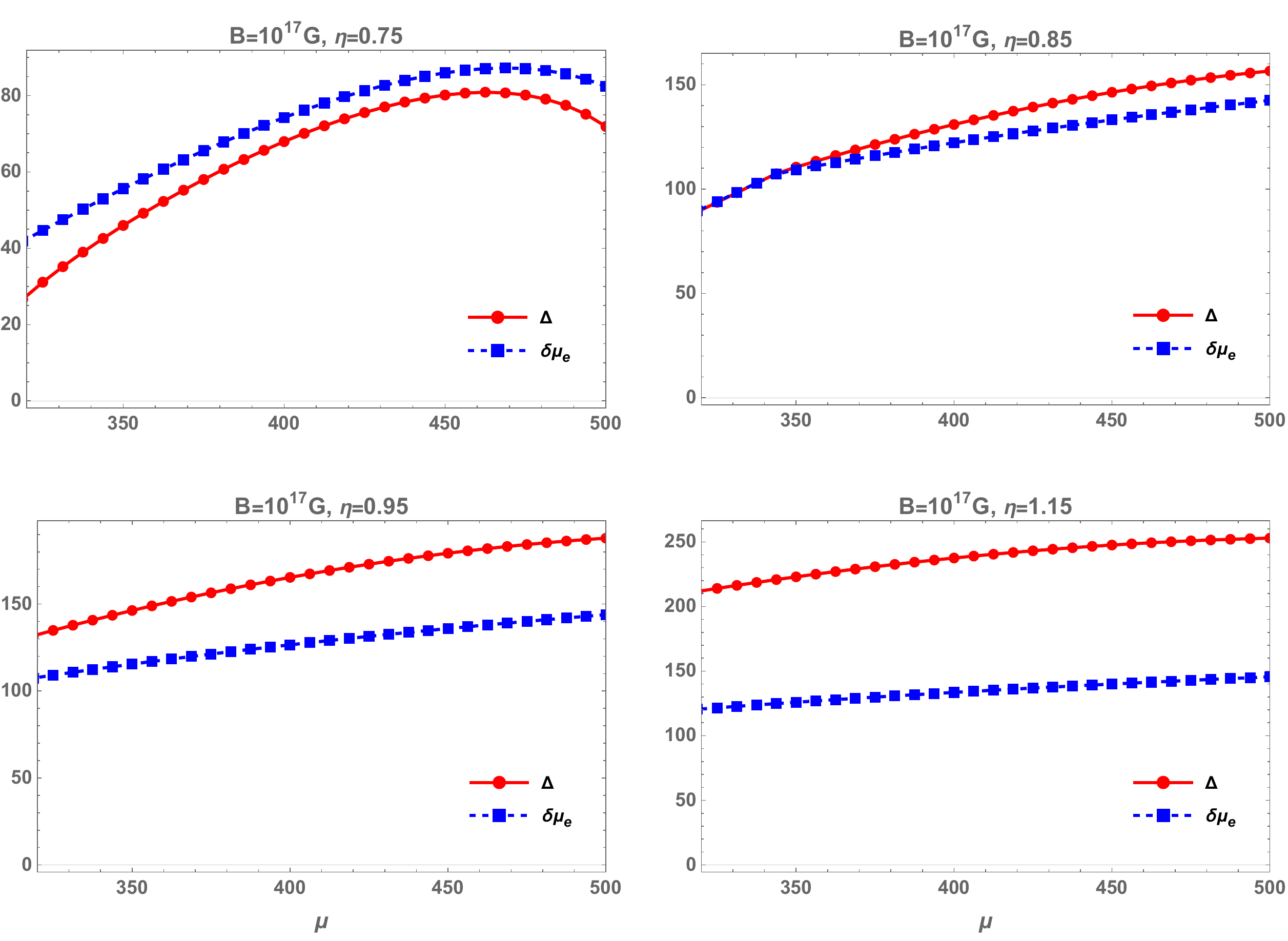}
	\caption{The gap $\Delta$ (in MeV) and mismatch $\delta \mu_e$ (in MeV) versus baryonic chemical potential $\mu$ (in MeV) for different diquark couplings at a magnetic field value of $10^{17}$G.} \label{Fig_1}
\end{figure}

\begin{figure}
	\includegraphics[height=11cm]{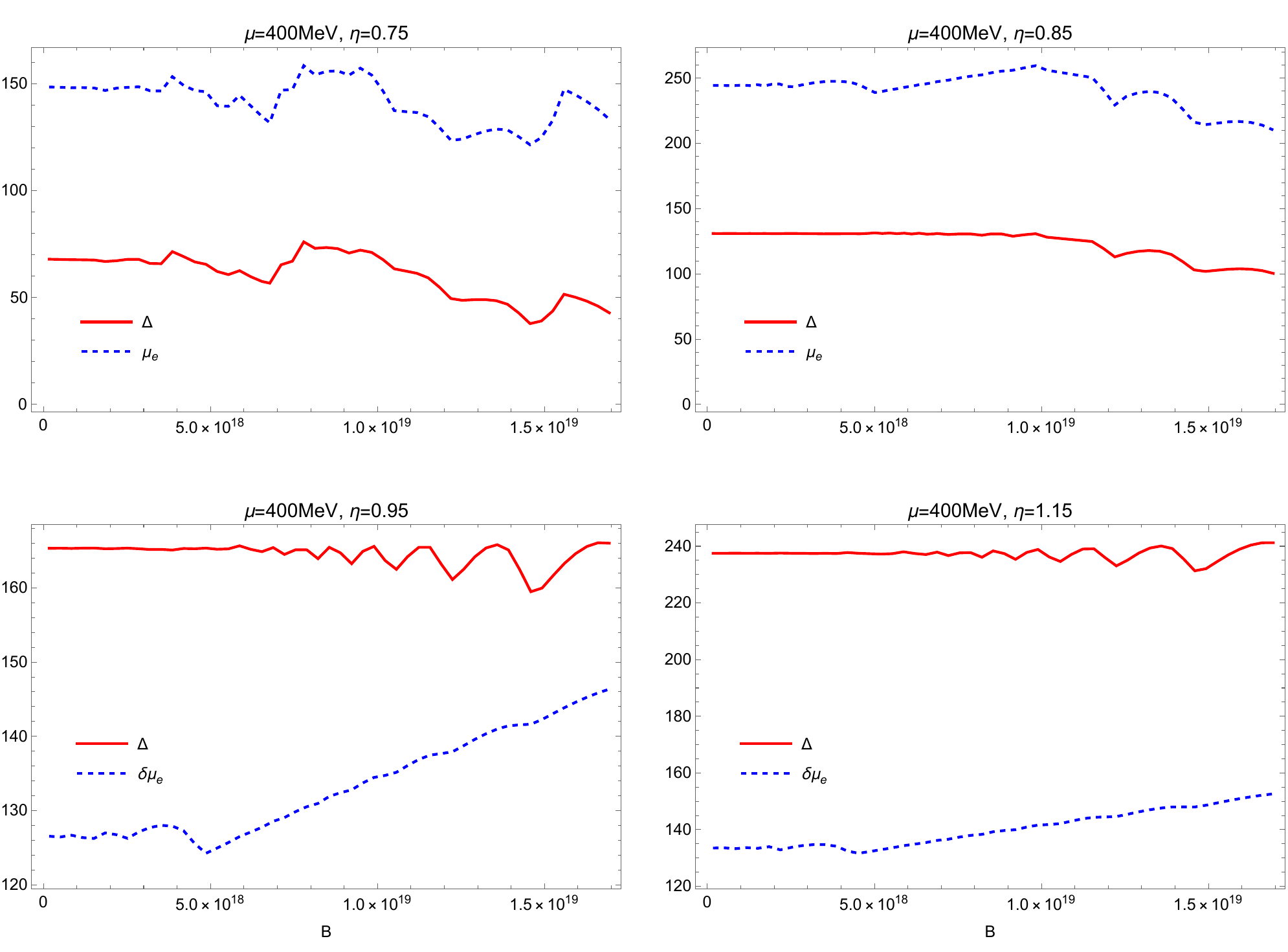}
	\caption{The gap $\Delta$ (in MeV) and mismatch $\delta \mu_e$ (in MeV) versus magnetic field $\tilde{B}$ (in Gauss) for different diquark couplings at baryonic chemical potential $\mu=400$ MeV.}\label{Fig_2}
\end{figure}

For a magnetic field of $10^{17}$ G, Fig.~\ref{Fig_1} shows the gap $\Delta$ and the mismatch $\delta\mu_e$ as functions of the baryonic chemical potential between $320$ MeV and $500$ MeV, for four different coupling strengths. We note that both $\Delta$ and $\delta\mu_e$ increase with increasing chemical potential indicating that as more quarks form Cooper pairs the system needs a larger amount of electrons to keep its electrical neutrality.  At a small coupling $\eta=0.75$, however,  $\Delta<\delta\mu_e$ and thus this is the g2SC phase, while at large couplings $\eta=0.95$ and $\eta=1.15$,  $\Delta>\delta\mu_e$ and it is reached the 2SC phase. The reason for this difference is due to the fact that $\Delta$ depends on the coupling strength more significantly than  $\delta\mu_e$ does as expected from the gap equations (\ref{gap-eq}). A detailed comparison of these results with that for magnetic field of value $10^{18} {\rm G}$ shows that the magnetic field does not significantly affect the behavior of $\Delta$ and $\delta\mu_e$ in the parameter range under consideration. 

For a fixed chemical potential $\mu=400$ MeV, Fig.~\ref{Fig_2} shows the gap $\Delta$ and the mismatch $\delta\mu_e$ as functions of the magnetic field for four different coupling strengths. Again we see the stronger the coupling strength is the larger $\Delta$ will be compared with $\delta\mu_e$. Both $\Delta$ and $\delta\mu_e$ remain at their zero field values up to a magnetic field as high as $10^{18} {\rm G}$, which justifies our claim that in the range of magnetic field of interest of NS both $\Delta$ and $\delta\mu_e$ do not change significantly. The so-called de Hass-van Alphen oscillation driven by the discreteness of Landau levels becomes noticeable at  higher value of magnetic field. These results are consistent with that obtained in previous studies \cite{Jaikumar-1, Jaikumar-2,Jaikumar-3,Scoccola, Mishra}.

\begin{figure}
	\includegraphics[height=10cm]{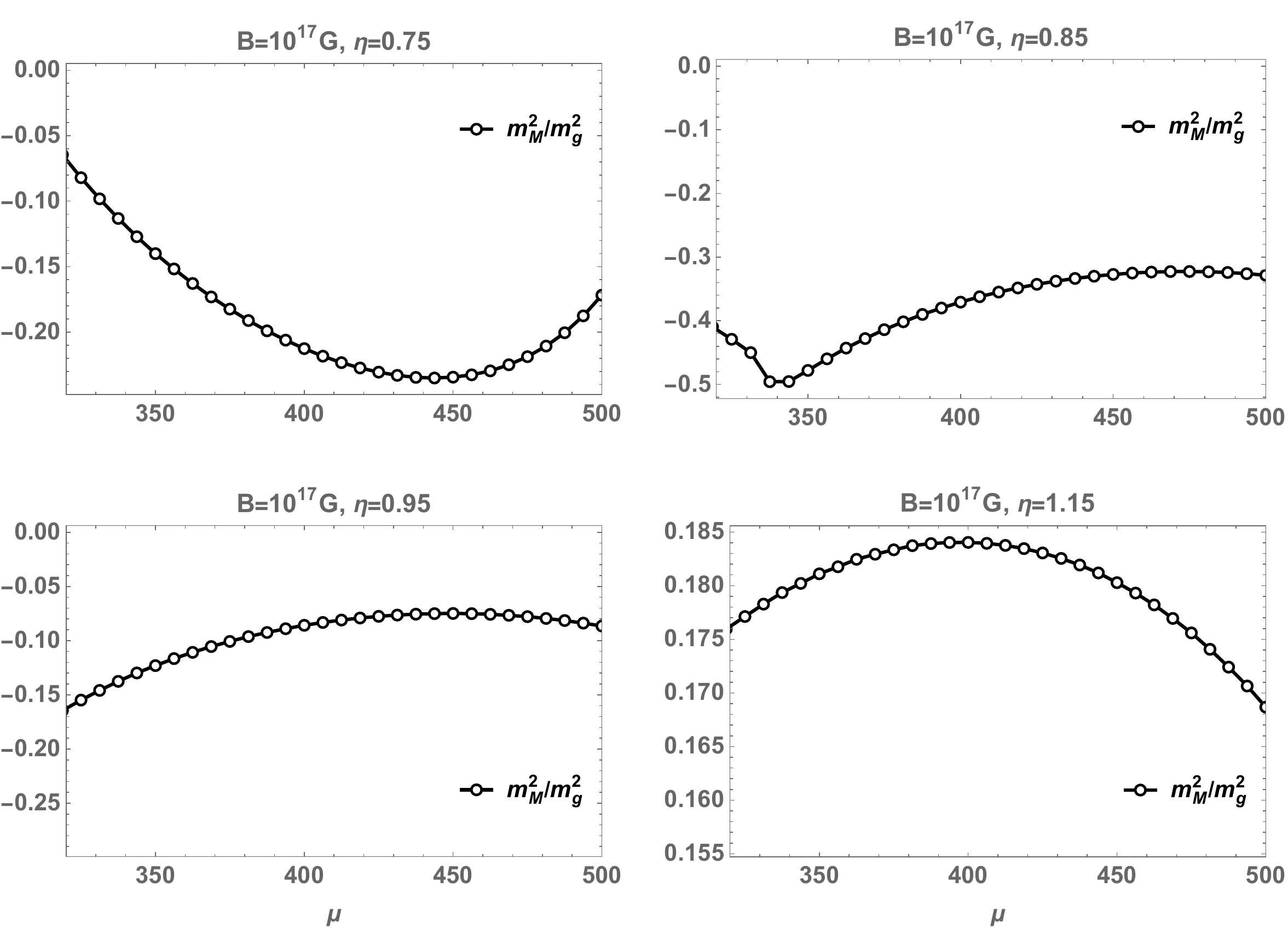}
	\caption{The ratio of the squared Meissner mass of charged gluons to the magnetic mass, $m_M^2/m_g^2$, versus the chemical potential $\mu$ (in MeV) for different diquark couplings at a magnetic field of $10^{17}$ G. } \label{Cromo-M-Inst-1}
\end{figure}


\subsection{Effect of strong magnetic field on the competition between chiral and diquark condensates}

In the previous section, we focused on the behavior of the diquark gap $\Delta$ and the electric chemical potential $\mu_e$ in the neutral 2SC phase, while neglecting the chiral condensate \cite{Igor-Rischke}. This is justified in the density regime of interest for neutron star interiors ($\mu \gtrsim 300$ MeV), where the chiral condensate essentially vanishes in the absence of a magnetic field. However, as we shall discuss below, a strong magnetic field can significantly alter the chiral condensate and its interplay with the diquark pairing, potentially modifying the phase structure of dense quark matter in magnetar cores.

The effect of a constant magnetic field on the color-neutral 2SC phase was systematically studied by Fayazbakhsh and Sadooghi \cite{Neutral-2SC-Magnetic-Field} using a two-flavor NJL model that includes both chiral and diquark condensates, as well as the color chemical potential $\mu_8$. They provided a detailed analytical and numerical survey of the gap equations across a wide range of magnetic field strengths. In the strong-field regime where the lowest Landau level (LLL) dominates, they derived analytic expressions for the chiral gap $\sigma$ and the diquark gap $\Delta$,
\begin{equation}
\sigma^2 \sim 4\tilde{e}\tilde{B} \, e^{-\frac{2\Lambda^2}{3\tilde{e}\tilde{B}}\left(\frac{1}{G_S}-1\right)},\qquad
\Delta^2 \sim 4(\Lambda_B^2-\mu^2)\, e^{-\frac{\Lambda^2}{\tilde{e}\tilde{B}}\frac{1}{G_D}}. \label{analytical-chiral-diquark-gap}
\end{equation}
Here $\Lambda$ is the three-momentum cutoff, while $\Lambda_B=\sqrt{\tilde{e}\tilde{B}}$
is the effective cutoff for the LLL contributions. Eqs. (\ref{analytical-chiral-diquark-gap}) explicitly demonstrate the magnetic catalysis effect, the enhancement of both condensates by the magnetic field.  For arbitrary magnetic fields, their numerical results reveal de Haas-van Alphen oscillations in the mass gaps and magnetizations at moderate fields, which transition into a linear LLL dominated regime for $\tilde{e}\tilde{B} \gtrsim 0.45$ GeV$^2$. They also mapped out the phase structure in the $\mu_c$-$\tilde{e}\tilde{B}$ plane, where \(\mu_c\) denotes the critical chemical potential at which the phase transition takes place, finding that the transition from the chiral symmetry breaking phase to the CS phase is of first order for $\mu_c \sim 350$--$450$ MeV across a broad range of magnetic fields. For nonvanishing $\mu_8$, they further showed that the second-order transition from the CS phase to the normal phase at high density can become first order.

Building upon this framework, the coupled evolution of the chiral condensate  and the diquark condensate was later investigated in detail by Mandal and Jaikumar \cite{Jaikumar-1, Jaikumar-2, Jaikumar-3}. Their study extends the earlier zero-field work \cite{Huang-Igor,Igor} and incorporates the effect of a strong homogeneous magnetic field on both order parameters in a self-consistent manner. The thermodynamic potential in the presence of a magnetic field is given by
\begin{equation}
\begin{aligned}
\Omega_{{\tilde B}\neq 0} =&\, \frac{\sigma^2}{4G_S} + \frac{\Delta^2}{4G_D} 
- \int_0^\infty \frac{d^3\mathbf{p}}{(2\pi)^3} \big[f(E_{p,0}^+) + f(E_{p,0}^-)\big] \\
& - \frac{\tilde{e}\tilde{B}}{8\pi^2} \sum_{n=0}^\infty \alpha_n \int_{-\infty}^\infty dp_z \Big[ f(E_{p,1}^+) + f(E_{p,1}^-) + 2 f(E_{\Delta,\frac{1}{2}}^+) + 2 f(E_{\Delta,\frac{1}{2}}^-) \Big],
\end{aligned}
\end{equation}
where $f(E) = E + 2T\ln(1+e^{-E/T})$ at finite temperature, and the quasi-particle energies $E_{p,a}$ and $E_{\Delta,a}$ are defined with the Landau-level quantization as in Eqs.~(\ref{eigenmodes}) of the present work. The gap equations are then obtained by minimizing $\Omega$ with respect to $m$ and $\Delta$:
\begin{equation}
\frac{\partial \Omega}{\partial m} = \frac{\partial \Omega}{\partial \Delta} = 0.
\end{equation}
Mandal and Jaikumar solved these coupled equations numerically for various values of the diquark coupling ratio $\rho \equiv G_D/G_S$ and magnetic field strengths. In the weak-field limit ($\tilde{B} \lesssim 5\times 10^{17}$ G), the number of occupied Landau levels is large, and the discrete sum approaches a continuum. In this regime, the results reproduce the zero-field behavior \cite{Igor, Huang-Igor-1}: for small $\rho$, both gaps exhibit a first-order transition; as $\rho$ increases beyond a critical value 
$\rho_c\simeq 1.09$, the first order transition softens into a smooth crossover, and a region emerges where both $m$ and $\Delta$
are nonzero, the so-called "mixed broken phase." However, a dramatic change occurs when the magnetic field is sufficiently strong ($\tilde{B} \gtrsim 5\times 10^{18}$ G). In this case, the mixed broken phase is completely destroyed, and the crossover observed at zero field is replaced by a first-order phase transition for both order parameters at essentially the same critical chemical potential. This is illustrated in Table II of Ref.~\cite{Jaikumar-3}, where for $\rho = 1.25$ the smooth crossover at $\mu_c^p \simeq 255$ MeV becomes a first-order transition with discontinuous jumps $\delta_m \simeq 185$ MeV and $\delta_\Delta \simeq 122$ MeV at $\mu_c \simeq 284$ MeV for $\tilde{e}\tilde{B} = 0.1$ GeV$^2$. The underlying physics can be understood as follows. A strong magnetic field acts as a stress on the chiral condensate, since it pairs quarks of opposite spin but same rotated charge, whereas it enhances the diquark condensate, which pairs quarks of opposite spin and opposite rotated charge. The modified Landau-level density of states thus strengthens the competition between the two condensates, leading to a sharpening of the transition and the disappearance of the coexistence region.

Taken together, these studies consistently indicate that while the diquark gap $\Delta$ remains robust in strong magnetic fields, the interplay between chiral and diquark condensates is significantly modified, with the magnetic field enhancing their competition and altering the nature of the phase transition. These findings are particularly relevant for the subsequent discussion of gluon screening and the vortex mechanism. First, in the strong field regime relevant to magnetar cores ($\tilde{B} \sim 10^{18}$--$10^{19}$ G), the diquark gap $\Delta$ may be accompanied by a non-negligible chiral gap $m$ near the transition region, which could affect the quasiparticle spectrum and hence the gluon polarization tensor. Second, the restoration of a first order transition in the presence of a strong magnetic field could lead to density discontinuities, and together with the constraints of electric and color neutrality, may induce mixed-phase domains in the stellar interior, potentially affecting the mechanical and thermal evolution of magnetars. Finally, and most relevant to the vortex scenario reviewed below, the fact that $\Delta$ remains robust in strong fields supports the viability of the 2SC phase as the background medium for the gluon-vortex instability. The values of $\Delta$ obtained from the coupled solution are of the same order as those used in Sec.~II.A, so the subsequent analysis of the Meissner masses and the field-boosting mechanism remains qualitatively valid. A quantitative refinement incorporating the magnetic-field dependence of $m$ into the gluon self-energy, however, remains an open question for future investigation.

\subsection{The  Chromomagnetic instability at ${\tilde{\bm B}}\ne 0$}


We now turn to the screening properties of gluon fields in the 2SC phase, which are directly relevant to the stability of the color superconducting ground state.  Meissner effect for the gluon fields arises in the 2SC phase due to the fact that the symmetry group $SU(3)_c\times U(1)_{\rm em}$ is broken down to $SU(2)_c\times {\tilde U}(1)_{\rm em}$ by the condensation of Cooper pairs \cite{Rischke-Debye-Meissner, Rischke-Shovkovy,Igor, Huang-Igor, Huang-Igor-1, Huang-Igor-2,Schmitt-Wang-Rischke}. In particular, the gluon screening properties in neutral 2SC phase had been considered in details in Ref.\cite{Huang-Igor-1,Huang-Igor-2} without the presence of a magnetic field.  It was found surprisingly that the Meissner masses of the four charged gluon fields $A^a_\mu$ for $a=4,5,6,7$ and the rotated 8th gluon field (a mixture of original 8th gluon and the photon) could be imaginary in some parameter regions.  To be explicit,
\begin{equation}
m^2_{M,{\rm charged}}=\frac{2\alpha_s{\bar\mu}^2}{3\pi}\left[\frac{\Delta^2-2(\delta\mu_e)^2}{\Delta^2}+2\frac{\delta\mu_e\sqrt{(\delta\mu_e)^2-\Delta^2}}{\Delta^2}\theta(\delta\mu_e-\Delta)\right],\label{4to7Meissner}
\end{equation}
and
\begin{equation}
m^2_{M,{\tilde 8}{\tilde 8}}=\frac{4(3\alpha_s+\alpha){\bar\mu}^2}{27\pi}\left[1-\frac{\delta\mu_e\theta(\delta\mu_e-\Delta)}{\sqrt{(\delta\mu_e)^2-\Delta^2}}\right],\label{8thMeissner}
\end{equation}
with $\alpha=e^2/4\pi$ and $\alpha_s=g^2/4\pi$.  The squared Meissner mass is negative for the charged gluons when $0<\Delta/\delta\mu_e<\sqrt{2}$ and  for the rotated  8th gluon when $0<\Delta/\delta\mu_e<1$. The existence of those negative squared Meissner masses signals an instability in the quark matter, which was named as the chromomagnetic instability.  It is worth noting that this instability even extends to the gapped 2SC phase ($1<\Delta/\delta\mu_e<\sqrt{2}$) as shown by Eq. (\ref{4to7Meissner}).   

The effect of a magnetic field on the gluon screening properties is two fold. On one hand, the dependence of the gap equations (\ref{gap-eq}) on the magnetic field renders the field dependence of the gap $\Delta$ and the mismatch $\delta\mu_e$ as shown in Fig.\ref{Fig_2}. Those dependence eventually enters into the Meissner masses (\ref{4to7Meissner}) and (\ref{8thMeissner}) though implicitly. On the other hand, the polarization tensor of charged gluons, from which the screening masses are derived, acquires a Lorentz structure that is no longer isotropic but exhibits an anisotropy along the field direction \cite{Pi_Anisotropy,Pi_Anisotropy-2,Bo-photon,Bo-Lack-Debye}. For neutron star applications, however, this anisotropy is only a subleading effect, since ${\tilde e}{\tilde B} \ll \mu^2$. Consequently, the polarization tensor can be approximated by its zero-field form for a dense medium, with the magnetic-field dependence entering only through $\Delta$ and $\mu_e$. It would nevertheless be worthwhile to study the charged-gluon polarization tensor in the regime ${\tilde e}{\tilde B} \sim \mu^2$, where the dimensional reduction of the quark loop becomes significant.

Fig. \ref{Cromo-M-Inst-1} displays the ratio of the charged-gluon Meissner mass to the gluon mass scale $m_g^2=4\alpha_s{\bar\mu}^2/3\pi$, plotted against the baryonic chemical potential relevant for NS. It can be seen that the Meissner mass remains tachyonic unless the coupling is very large ($\eta=1.15$). The reason is that for sufficiently strong coupling the condition $\Delta/\delta \mu_e>\sqrt{2}$ is fulfilled, as illustrated in the last panel of Fig.~\ref{Fig_1}. We also find that at strong fields around $10^{18}$ G, $m_M^2$ stays essentially unchanged, as expected. Finally, we emphasize that $m_M^2$ does not vary monotonically with the baryonic chemical potential, because it depends on two quantities, $\Delta$ and $\delta \mu_e$, which have different dependences on $\mu$.

\section{The zero-mode problem and gluon condensate in neutral color superconductiviting phase}

\subsection{Zero-mode problem}
To some extend, the chromomagnetic instability in magnetized neutral CS phase is connected to the well-known ``zero-mode problem" associated with the interplay between charged Yang-Mills fields and a magnetic field \cite{MCFL-Vortices,Ferrer-Vortices1,MCFL-Vortices-2,Yang-Mills,Yang-Mills_2,W-EW1,W-EW2,W-EW-2,W-EW-3,String Theory}. In this section, we will give a brief review of the ``zero-mode problem" and the resulting gluon condensate effect.  The effective action for the gauge fields including the one-loop quark contribution can be written as
\begin{align}
\nonumber \Gamma_{\rm eff}=&-\frac{1}{4}\int d^4x\left[(G^a_{\mu\nu})^2+(f_{\mu\nu})^2\right]-\frac{1}{2}\int d^4xd^4yG^a_\mu(x)\Pi^{ab}_{\mu\nu}(x,y)G^b_\nu(y)-\frac{1}{2}\int d^4xd^4yA_\mu(x)\Pi_{\mu\nu}(x,y)A_\nu(y)\\
&+\int d^4x {\cal L}_{\rm gf}(x)\label{effaction}
\end{align} 
where $G^a_{\mu\nu}$ and $f_{\mu\nu}$ are the gluon and electromagnetic field strength tensors, and $\Pi^{ab}_{\mu\nu}$ and $\Pi_{\mu\nu}$ are their corresponding polarization operators, respectively. ${\cal L}_{\rm gf}$ is the gauge fixing Lagrangian density. Specific to the charged gluon fields in CS phases, say $G_\mu^\pm$ in Table \ref{chargedgluons}, whose polarization tensor in the infrared limit can be written as: $\Pi^{ab}_{\mu\nu}=\delta^{ab}\delta(x-y)\left(m_D^2\delta_{\mu 0}\delta_{\nu 0}+m^2_M\delta_{\mu i}\delta_{\nu i}\right)$ with $m_D$ and $m_M$ the Debye and Meissner mass respectively. Then the effective action (\ref{effaction}) for $G^\pm$ in the background of an applied rotated magnetic field takes the form
\begin{align}
\nonumber \Gamma_{\rm eff}=&\int d^4x \left[-\frac{1}{4}({\tilde f}_{\mu\nu})^2-\frac{1}{2}\left({\tilde\Pi}_\mu G_\nu^--{\tilde\Pi}_\nu G_\mu^-\right)^2-\left(m_M^2\delta_{\mu i}\delta_{\nu i}+i{\tilde e}{\tilde f}_{\mu\nu} \right)G_\mu^+G_\nu^-\right.\\
&+\left.\frac{g^2}{2}\left[(G_\mu^+)^2(G_\nu^-)^2-(G_\mu^+G_\nu^-)^2 \right]+\frac{1}{\lambda}G_\mu^+{\tilde\Pi}_\mu{\tilde\Pi}_\nu G_\nu^-\right],\label{effectiveaction}
\end{align}
with ${\tilde\Pi}_\mu=\partial_\mu-i{\tilde e}{\tilde A}_\mu$  and the covariant derivative, $\tilde{f}_{\mu \nu}=\partial_\mu \tilde{A}_\nu-\partial_\nu \tilde{A}_\mu$. We take the external rotated magnetic field along $z$ direction and thus the nonvanishing component is ${\tilde f}_{12}\equiv {\tilde H}$. In (\ref{effectiveaction}), the last term comes from the gauge fixing term in (\ref{effaction}) taken in the `t Hooft gauge with arbitrary gauge fixing paramter $\lambda$. Moreover, we ignore the Debye mass $m_D$ since it will not be substantial for our discussions. The presence of an anomalous magnetic moment term $i{\tilde e}{\tilde f}_{\mu\nu} G_\mu^+G_\nu^-$ can render a gluon mode tachyonic when the magnetic field exceeds a critical value, even if $m_M^2$ is initially positive.  This can be illustrated by diagonalizing the mass matrix of the field components $G_{1,2}^\pm$
\begin{equation}
\left(\begin{array}{cc}
m_M^2 & i{\tilde e}{\tilde H}\\
-i{\tilde e}{\tilde H} & m_M^2
\end{array}\right)\longrightarrow \left(\begin{array}{cc}
m_M^2+{\tilde e}{\tilde H} & 0\\
0 & m_M^2-{\tilde e}{\tilde H}
\end{array}\right)\label{diagonalization}
\end{equation}
Obviously, once the magnetic field is beyong the threshold ${\tilde e}{\tilde H}_C=m_M^2$ one of the mode in (\ref{diagonalization}) becomes tachyonic. This is the well known``zero-mode problem" for the Yang-Mills fields in the presence of a magnetic field \cite{Yang-Mills,Yang-Mills_2}, for the $W^\pm_\mu$ bosons in the electroweak theory \cite{W-EW1,W-EW2,W-EW-2,W-EW-3}, and for higher-spin fields in the context of string theory \cite{String Theory}.

\begin{figure}
	\includegraphics[height=6cm]{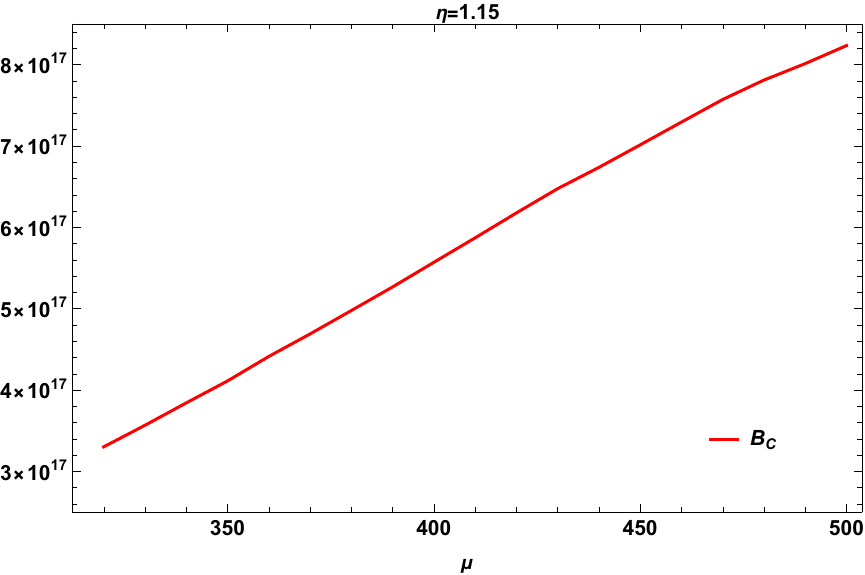}
	\caption{The critical field ${\tilde B}_c$ (in Gauss) in terms of the chemical potential $\mu$ (in MeV) for the coupling strength $\eta=1.15$.} \label{Critical_field}
\end{figure}

The solution to this zero-mode problem is the spontaneous generation of gluon field condensates $\langle G_\mu^\pm\rangle$. Following the ansatz employed in Ref.  \cite{MCFL-Vortices} where $\langle G_1^-\rangle=-i\langle G_2^-\rangle= G(x,y)$, $\langle G_3^-\rangle=\langle G_0^-\rangle=0$ and $\langle G_i^+\rangle=\langle G_i^-\rangle^*$,  one can find the Gibbs free energy
\begin{equation}\label{E-A-1}
{\cal G}={\cal F}_{0}-2 G^\dagger{\tilde\Pi}^2 G-2(2{\tilde e}{\tilde B}-m_M^2)|G|^2+2g^2|G|^4+\frac{1}{2}{\tilde B}^2-{\tilde H}{\tilde B},
\end{equation}
where $\tilde B$ is the in-medium induced magnetic fields and ${\cal F}_0$ is the free energy density of the normal phase ($G=0$) at zero applied field. As can be seen from the third term of the RHS of (\ref{E-A-1}) ${\tilde e}\tilde{B}$ enters as a negative square mass. In the case that $m_M^2 > 0$,  the ground state of the system has to be modified when ${\tilde e}\tilde{B} >m_M^2/2$. In Fig.\ref{Critical_field}, we show the critical field in terms of the baryonic chemical potential for the coupling strength $\eta=1.15$ in neutral 2SC matter, which is strong enough to validate the condition $\Delta/\delta\mu_e>\sqrt{2}$ and thus $m_M^2>0$.  Throughout the range of chemical potential of interest in NS, the magnitude of the critical field is of the order of $10^{17}$ G. In our present case, however, we are only interested in the region of moderate values of baryonic density and coupling strength that makes $m_M^2 < 0$, which is even more  pressing since a magnetic field as weak as possible will activate the mechanism (i.e. no critical field is needed).


To find the new ground state where the instability is removed, the Gibbs free energy should be minimized with respect to $G$ and $\tilde{B}$. Hence, we find the set of equations
\begin{equation}
-{\tilde\Pi}^2G-(2{\tilde e}{\tilde B}-m_M^2)G+2g^2|G|^2G=0,\label{gluoncondensateequation}
\end{equation}
and
\begin{equation}
2{\tilde e}|G|^2-{\tilde B}+{\tilde H}=0, \label{inducedmagneticfield}
\end{equation}
from which the condensate solution $G$ and the induced magnetic field $\tilde B$ are solved. For $m_M^2>0$ and an applied field near the critical value ${\tilde H}_c$, it was found \cite{MCFL-Vortices} that the explicit solution of (\ref{gluoncondensateequation}) with vanishing conditions at $x\rightarrow \pm\infty$ is
\begin{equation}
G_k=\exp(-ik_yy)\exp\left[-\frac{(x-x_k)^2}{2\xi^2}\right],\label{gluonvortex}
\end{equation}
with $x_k\equiv k_y/{\tilde e}{\tilde H}_c$ and $\xi^2\equiv 1/(2{\tilde e}{\tilde H}_c-m_M^2)$. The general solution $G(x,y)$ is the superposition of all the Gaussian solutions (\ref{gluonvortex}) centered at different $x_k$, which constitutes the vortex state that removes the instability in the whole space. 

\subsection{Gluonic phase as a general framework for gluon condensation}

The vortex solution discussed above is one concrete realization of a more general mechanism in which gluonic degrees of freedom play an active role in restructuring the ground state of neutral 2SC matter \cite{Gluonic_phase, Gluonic_phase_1, Gluonic_phase_2, Gluonic_phase_review}. A systematic study of this phenomenon was carried out by Gorbar, Hashimoto, and Miransky using a Ginzburg-Landau (GL) approach within a gauged Nambu-Jona-Lasinio model \cite{Gluonic_phase}. They showed that the chromomagnetic instability of the 2SC phase can be cured by the formation of vector condensates of gluons, leading to a new phase termed as the "gluonic phase".

Near the critical point $\delta \mu = \Delta/\sqrt{2}$, where the Meissner mass of the charged gluons vanishes, the relevant light degrees of freedom are the gluons of the unbroken $SU(2)_c$ subgroup and the complex doublet $K_\mu$ formed by $A_\mu^{(4)-(7)}$. The effective GL potential for the vector condensates was derived as
\begin{equation}
V_{\rm eff} = V_\Delta + \frac{1}{2} M_B^2 B^2 + T D B^2 + \frac{1}{2}\lambda_{BC} B^2 C^2 + \frac{1}{2}\lambda_{CD} C^2 D^2,
\end{equation}
where $V_\Delta$ is the 2SC part of the effective potential and the dimensionless coefficients $B \equiv g\langle A_3^{(6)}\rangle$, $C \equiv g\langle A_3^{(1)}\rangle$, and $D \equiv g\langle A_3^{(1)}\rangle$ denote the gluon vector condensates along the third spatial direction. The coefficient $M_B^2$ is directly related to the gluon Meissner mass and becomes negative when the 2SC solution turns unstable. The quartic couplings $\lambda_{BC}>0$ and $\lambda_{CD}<0$ characterize the self-interactions among the condensates. Most importantly, the coefficient $T$ is a dimensionful (mass dimension one) trilinear coupling that arises from both the tree-level gluon self-interactions and the one-loop quark contributions. Its presence in the term $TDB^2$ is crucial because it induces anomalous scaling relations for the condensates near the critical point, as we shall see below, and ensures that the gluonic phase is energetically favored over the 2SC state. The stationary solution of the potential yields
\begin{equation}
B_{\rm sol} \propto -M_B^2,\qquad C_{\rm sol} \propto \sqrt{-M_B^2},\qquad D_{\rm sol} \propto -M_B^2,
\end{equation}
and the corresponding free energy is lower than that of the 2SC phase, confirming that the gluonic phase is indeed more stable.

A crucial feature of this phase is that the vector condensates spontaneously break the color $SU(2)_c$, the electromagnetic $\tilde{U}(1)_{\rm em}$, and the rotational $SO(3)_{\rm rot}$ symmetries, leaving only a residual $SO(2)$ rotational symmetry. The symmetry breaking pattern is
\begin{equation}
SU(3)_c \times U(1)_{\rm em} \times SO(3)_{\rm rot}
\;\xrightarrow{\Delta}\;
SU(2)_c \times \tilde{U}(1)_{\rm em} \times SO(3)_{\rm rot}
\;\xrightarrow{\langle K_3\rangle}\;
\tilde{U}(1)_{\rm em} \times SO(2)_{\rm rot},
\end{equation}
so the gluonic phase describes an anisotropic medium in which color and electric superconductivities coexist. Furthermore, the condensates generate non-Abelian constant chromoelectric fields, e.g., $E_3^{(2)} = C_{\rm sol}D_{\rm sol}/g$ and $E_3^{(7)} = B_{\rm sol}(2\mu_8 - D_{\rm sol})/(2g)$, which do not produce instabilities because they are expressed through constant vector potentials, preserving momentum and energy as good quantum numbers.


The existence of topologically stable vortex solutions in theories with vector condensates sharing the same symmetry breaking pattern as the gluonic phase has been demonstrated in a simplified gauged SU(2)$\times$U(1)$_Y$ $\sigma$-model with a hypercharge chemical potential, where magnetic, hypermagnetic, and hybrid vortices were found \cite{Gluonic_phase_3}.

It is instructive to compare the gluonic phase with the inhomogeneous gluon vortex condensate discussed in Sec.~III.A. This issue was addressed in \cite{Ferrer-Vortices1}, in which the free energies of the homogeneous gluonic phase and the inhomogeneous gluon vortex condensate with an induced magnetic field were compared in the context of the neutral 2SC phase. They found that near the critical point $\delta \mu \simeq \Delta/\sqrt{2}$, where the squared Meissner mass $m_M^2$ is small, the homogeneous gluon condensate proposed in the gluonic phase yields a condensation energy density of order
\begin{equation}
\overline{\mathcal{F}}_g^{\rm hom} \simeq - \frac{\pi}{2\alpha_s^3}\frac{(-m_M^2)^3}{m_g^2},
\end{equation}
with $m_g^2 = 4\alpha_s^2 \bar{\mu}^2/(3\pi)$ being the gluon magnetic mass scale. In contrast, the inhomogeneous condensate $\langle K_i\rangle$ that induces a rotated magnetic field $\tilde{B}$ gives a condensation energy density
\begin{equation}
\overline{\mathcal{F}}_g^{\rm inh} \simeq - \frac{c^2 \pi}{2} \frac{-m_M^2}{\tilde{q}^2},
\end{equation}
where $c$ is a dimensionless constant characterizing the vortex cell size, and $\tilde{q}$ is the rotated charge of the gluon field. Since near the transition point $|m_M^2| \ll m_g^2$  in the weak coupling regime, the inhomogeneous vortex condensate is energetically favored over the homogeneous gluonic phase. This result suggests that, in the presence of a magnetic field,  the actual ground state of the neutral 2SC matter in the chromomagnetically unstable region is more likely to be an inhomogeneous gluon vortex condensate accompanied by an induced magnetic field, rather than a spatially homogeneous gluon condensate.

\section{The boosting mechanism and possible origin of the inner magnetic field of magnetars}

Having established that the gluon vortex condensate resolves the chromomagnetic instability, we now examine its most notable consequence: the paramagnetic amplification of the in-medium magnetic field. Note that in (\ref{inducedmagneticfield}) because of the different sign in the applied and induced field, the field $\tilde B$ can be dynamically induced by the contribution of the term $2{\tilde e}|G|^2$, which comes from the gluon anomalous magnetic moment $i{\tilde e}{\tilde f}_{\mu\nu} G_\mu^+G_\nu^-$ in (\ref{effectiveaction})  making $\tilde{B}$ larger than the applied field $\tilde H$. Thus, once a gluon condensate develops, the magnetic field will be antiscreened and the neutral 2SC phase will present paramagnetism. This is the magnetic field boosting mechanism that been reported in \cite{BF-Ef}.
In order to examine quantitatively the magnetic field boosting effect in neutral 2SC phase, we first transform (\ref{gluoncondensateequation}) and  (\ref{inducedmagneticfield}) to momentum space and obtain
\begin{equation}
\left\{-\left[(p^0)^2-{\tilde e}{\tilde B}(2n+1) \right]-(2{\tilde e}{\tilde B}-m_M^2)-\frac{g^2({\tilde H}-{\tilde B})}{{\tilde e}}   \right\}G(p)=0.
\end{equation}
We can then deduce the dispersion relation of the charged gluons
\begin{equation}
E_n^2={\tilde e}{\tilde B}(2n+1)-(2{\tilde e}{\tilde B}-m_M^2)-\frac{g^2({\tilde H}-{\tilde B})}{{\tilde e}}. 
\end{equation}
For the system to remain stable, the lowest energy mode of the charged gluons must satisfy
\begin{equation}
m_M^2-{\tilde e}{\tilde B}-\frac{g^2({\tilde H}-{\tilde B})}{{\tilde e}}\leq(2\Delta)^2,
\end{equation}
ensuring that no mode exceeds twice the quasiparticle gap. Therefore, the equality 
\begin{equation}
m_M^2-{\tilde e}{\tilde B}-\frac{g^2({\tilde H}-{\tilde B})}{{\tilde e}}=4\Delta^2,\label{inducedfield}
\end{equation}
will give the induced field $\tilde B$ for each applied field $\tilde H$. From the solutions of the gap equation and the $\tilde H$ field dependent squared Meissner mass given in Figs. \ref{Fig_1}-\ref{Cromo-M-Inst-1}, together with Eq.(\ref{inducedfield}), one can find the induced field $\tilde{B}$ as a function of the applied field $\tilde{H}$. 


By numerically solving Eq. (\ref{inducedfield}) with the solutions of the gap equation and the field-dependent squared Meissner mass given in Figs. 1-3, we obtain the induced field $\tilde{B}$
as a function of the applied field $\tilde{H}$ \cite{BF-Ef}. For a coupling strength 
$\eta=0.85$ and baryonic chemical potential $\mu=400 $ MeV, the boosting effect is most pronounced at applied fields around $10^{15}$ G, where the induced field is enhanced by up to two orders of magnitude. In this regime, the main contributions come from $m_M^2$ and $\Delta^2$. As the applied field increases toward $10^{18}$ G, the induced field and the applied field become comparable in magnitude, indicating that the contributions from $m_M^2$ and $\Delta^2$ become negligible. This behavior is expected because for the baryonic chemical potential under consideration, both applied and induced field strengths are weak compared to $\mu^2$, so the gap $\Delta$ and mismatch $\delta\mu_e$ remain close to their zero-field values. Consequently, decreasing the applied magnetic field below $10^{15}$ G does not significantly change the induced field from the value obtained at $10^{15}$ G.

To assess the viability of the above microscopic mechanism for magnetar applications, it is useful to compare it with the conventional scenario for magnetic field generation in neutron stars. One such scenario is the dynamo mechanism \cite{Thompson, Dynamo}, which converts the kinetic energy of an electrically conducting fluid into magnetic energy. Based on this conversion, we are able to obtain
\begin{equation}\label{EPE-Sun}
B=B_{\odot} \sqrt{\frac{MR_{\odot}}{M_{\odot} R}} \frac {P_{\odot}}{P}=\left ( \frac{0.8 \times 10^9} {P}  \right ) \textrm{G}, 
\end{equation}
with $M_{\odot}$, $R_{\odot}$ and $P_{\odot}$ being the solar mass, radius and rotational period respectively.
Since the Sun is not a rigid body, we will consider here its minimum value of 25.67 days that corresponds to  $P_{\odot}=2.2\times 10^6$ s and $B_\odot=1$ G. Taking the accepted values $M=1.4 M_{\odot}$ and $R=(0.14\times 10^{-4})R_\odot$, and using the spin periods $P$ from the McGill Online Magnetar Catalog \cite{McGill}, Eq.~(\ref{EPE-Sun}) yields magnetic field strengths on the order of $10^{15}$ G,  which is far from the expected $10^{17}$ G inner field of magnetars.

Hence, we see that the order of magnitude of the magnetic field which corresponds to the macroscopic rotational kinetic energy of the star is not enough to induce the large inner field of the magnetars. This result is suggesting \cite{BF-Ef} that the origin of this field can be associated with a microscopic mechanism that should be able to boost the inner field to larger values. The magnetic field boosting mechanism found in \cite{BF-Ef} can bridge the gap. As shown by the numerical results in the previous section, when the stellar core consists of neutral 2SC matter in a chromomagnetically unstable state, an external field around $10^{15}$ G can be amplified to approximately $10^{17}$ G, consistent with theoretical expectations for magnetar interiors.

We note that the simple rotational energy scaling of Eq. (\ref{EPE-Sun}) represents only a macroscopic estimate based on the star’s rotational kinetic energy alone. It does not account for other energy sources available in the early proto-neutron star (PNS) phase. Modern PNS dynamo simulations, including convective dynamos \cite{Raynaud2020,Matsumoto2022}, the Tayler-Spruit dynamo \cite{Barrere2025}, and MRI-driven $\alpha\Omega$ dynamos \cite{ReboulSalze2021,ReboulSalze2022}, consistently generate large scale fields of order $10^{15}$ G and toroidal fields reaching $10^{16}$ G in rapidly rotating proto-neutron stars.
The field amplification mechanism reviewed here is complementary to these macroscopic dynamos.  Once a seed field of order $10^{15}\sim 10^{16}$ G is established by the PNS dynamo, the gluon vortex condensate can further amplify it to $\sim 10^{17}$ G, as demonstrated in Fig.5. This two stage picture, macroscopic dynamo generating $10^{15}\sim 10^{16}$ G and followed by microscopic core boosting to $\sim 10^{17}$ G, provides a more complete and robust explanation for the ultra-strong inner fields of magnetars.

Remarkably, this boosting mechanism not only works for moderately large applied magnetic field but also for fields with smaller magnitudes.  Starting from (\ref{inducedfield}) 
\begin{equation}
{\tilde e}{\tilde B}\left(\frac{g^2}{{{\tilde e}}^2}-1\right)=\frac{g^2}{{{\tilde e}}^2}{\tilde e}{\tilde H}+4\Delta^2-m_M^2,
\end{equation}
in the regime of weak applied fields, where $\frac{g^2}{{{\tilde e}}}{\tilde H}$ is negligible compared to $4\Delta^2-m_M^2$ and $m_M^2$ is negative, the induced field approximately satisfies
\begin{equation}
{\tilde e}{\tilde B}\left(\frac{g^2}{{{\tilde e}}^2}-1\right)\simeq 4\Delta^2-m_M^2.\label{approximateboostedfield}
\end{equation}
Thus, if the diquark coupling strength is not so large that makes the squared Meissner mass positive and comparable with $4\Delta^2$, the induced magnetic field  $\tilde B$ becomes of the order of the larger parameter between $\Delta$ and $m_M$.
Therefore the dependence of the boosted field on the external field exists only implicitly in the gap and Meissner mass, which are almost constant for small and moderate external fields as shown in Fig.2. Hence, the magnitude of an induced field $\tilde{B}$ for an applied field $\tilde{H}$ of order of $10^{8}\sim 10^9$ G will be essentially the same as that for $\tilde{H}$ of order of $10^{15}$ G.  We therefore conclude that in the region with $\mu=400$ MeV and $\eta \in [0.75, 0.95]$ the induced field is of order $\tilde {B} \sim 10^{17}$ G, which is the expected value for the inner field of magnetars.

\begin{figure}\label{Boosted_field_for_different_eta}
	\includegraphics[height=6cm]{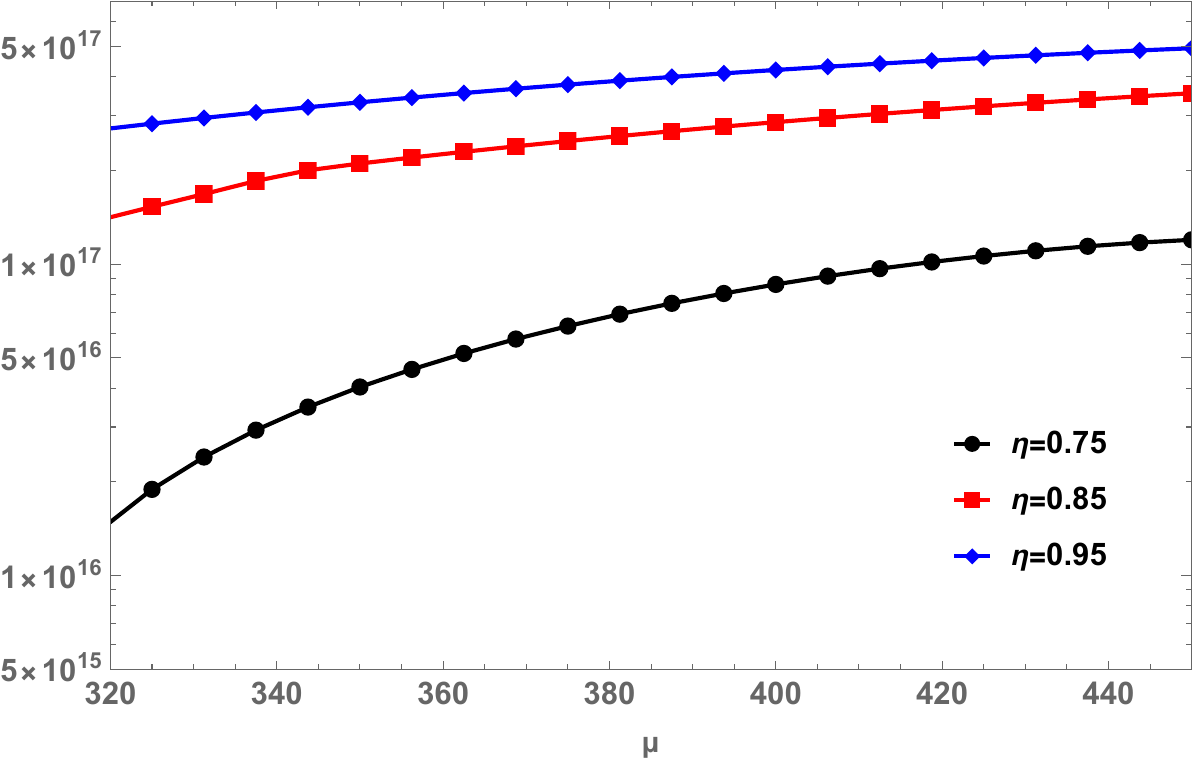} \caption{The induced field $\tilde{B}$ (in Gauss) versus the chemical potential $\mu$ (in MeV) for different diquark couplings \cite{BF-Ef}. The curves from bottom to top correspond to $\eta = 0.75, 0.85$, and $0.95$, respectively.}
\end{figure}

\section{Concluding remarks}

In this paper, we have reviewed the effect of an applied magnetic field on neutral dense quark matter, with particular emphasis on its implications for magnetar interiors. Using a two-flavor Nambu-Jona-Lasinio model in the mean-field approximation, we first solved the gap equations for the diquark condensate $\Delta$ and the electric chemical potential $\mu_e$ under a uniform external magnetic field. Our results show that for magnetic field strengths up to $10^{18}$ G, both $\Delta$ and $\mu_e$ remain largely unaffected, exhibiting only mild oscillations at the highest field values due to Landau-level discretization. We then extended this analysis to include the coupled evolution of the chiral and diquark order parameters\cite{Neutral-2SC-Magnetic-Field, Jaikumar-1, Jaikumar-2,Jaikumar-3, Scoccola, Mishra}. In the weak-field limit, the results reproduce the zero-field behavior: a smooth crossover emerges for sufficiently large diquark coupling, with a region where both condensates coexist. However, for strong magnetic fields $B \gtrsim 10^{18}$ G, the mixed broken phase is destroyed and the crossover is replaced by a first-order transition for both order parameters at essentially the same critical chemical potential. This finding is particularly relevant for magnetar cores, where such strong fields are expected, and it underscores that the diquark gap $\Delta$ remains robust in the strong-field regime, supporting the viability of the 2SC phase as the background medium for the gluonic instability discussed later.

We then computed the Meissner masses of the rotationally charged gluon fields and demonstrated that for moderate diquark couplings, the squared Meissner mass becomes negative across the entire density range relevant to neutron star interiors. This tachyonic behavior signals the persistence of the chromomagnetic instability even in the presence of a magnetic field. The presence of a rotated magnetic field can even enlarge the parameter region where the instability occurs due to the zero-mode problem. We showed that this instability can be resolved by restructuring the ground state through the formation of a gluon vortex condensate. The zero-mode problem can be understood from the effective action for charged gluons in a magnetic field, which contains an anomalous magnetic moment term $i\tilde{e}\tilde{f}_{\mu\nu}G_\mu^+ G_\nu^-$ that introduces a negative contribution to the squared mass. When this contribution dominates, the instability is inevitably triggered. In the case of neutral quark matter with $m_M^2<0$, the zero-mode effect is even more dramatic. A more general Ginzburg-Landau framework for gluon condensation was also reviewed: the gluonic phase proposed in Refs.~\cite{Gluonic_phase, Gluonic_phase_1, Gluonic_phase_2, Gluonic_phase_review, Gluonic_phase_3}. This phase features vector condensates of gluons that spontaneously break the color \(SU(2)_c\), the electromagnetic \(\tilde{U}(1)_{\rm em}\), and the rotational \(SO(3)_{\rm rot}\) symmetries, thereby producing an anisotropic medium with coexisting color and electric superconductivities. This framework is closely related to the vortex condensate scenario and provides a broader theoretical context for understanding gluonic degrees of freedom in neutral 2SC matter.

The key point of this review is that the gluon vortex condensate not only stabilizes the system but also induces a paramagnetic response that significantly amplifies the in-medium magnetic field. Solving the coupled equations for the condensate and the induced field, it was found \cite{BF-Ef} that seed fields on the order of $10^{15}$ G can be boosted by up to two orders of magnitude, reaching values of $10^{17}$ G precisely the range predicted for magnetar cores. Remarkably, this boosting mechanism operates even for much weaker applied fields, as the induced field is primarily determined by the gap and Meissner mass, which remain nearly constant for moderate external fields. Thus, fields as low as $10^8\sim 10^9$ G can be amplified to the same $10^{17}$ G scale. The dependence of the induced field on the baryonic chemical potential and diquark coupling strength was also examined, showing that the boosting effect is most pronounced in the moderate coupling regime relevant to compact star interiors.

Comparing this mechanism with the conventional dynamo explanation for magnetar fields, we show that while the dynamo mechanism can account for fields up to $10^{15}$ G by converting rotational kinetic energy into magnetic energy, it falls short of producing the $10^{17}$ G fields expected in magnetar cores \cite{Uryu}. The proposed microscopic mechanism naturally bridges this gap. Hence, this scenario suggests that if the inner core of a magnetar is composed of neutral 2SC quark matter in the chromomagnetically unstable regime, the resulting gluon condensation can provide the necessary field amplification.
Although the present analysis is based on an effective model of QCD rather than on first-principles QCD, and the diquark coupling $G_D$ is model dependent, the boosting mechanism is robust in the phenomenologically relevant range 
$\eta=0.75\sim 0.95$: for all these couplings the induced field reaches the order $10^{17}$ G, independently of the precise value of $G_D$.

We emphasize that the gluon-vortex mechanism discussed here is conceptually independent of the NJL framework. The NJL model is used only to estimate the quark-loop contribution to the gluon polarization, while the vortex/gluon-condensate picture itself follows from the general zero-mode and Ginzburg--Landau description. We also note that quark loops are known to modify the effective coupling \(G_S\) significantly \cite{EPJA2022}; therefore, a more robust phenomenological treatment should treat the diquark coupling independently of \(G_S\), rather than relying on the simple convention \(G_D = \eta G_S\) adopted here.

Several aspects of this review merit further investigation. First, the present analysis assumed a uniform background magnetic field and neglected possible anisotropies in the polarization tensor that could become important for extremely strong fields ($\tilde{e}\tilde{B} \sim \mu^2$). A more complete treatment incorporating the full tensorial structure of the gluon self-energy would be desirable. Second, the role of the strange quark and the possibility of transitioning to the CFL phase at higher densities remain to be explored \cite{MCFL-Vortices, Ferrer-Vortices1}. Third, a more systematic study of the model dependence of our results is desirable. This could be achieved by improving the effective model treatment, for instance by including vector repulsion \cite{vector_repulsive_1, vector_repulsive_2} and a running diquark coupling \cite{NJL_running_coupling_1, NJL_running_coupling_2}, so as to further test the robustness of the field amplification mechanism. Finally, connecting our mechanism to observable signatures, such as effects on the equation of state, cooling rates, and gravitational wave emission from neutron star mergers, would provide valuable links to astrophysical observations.

In conclusion, we have reviewed a novel mechanism for magnetic field amplification in dense quark matter that arises naturally from the chromomagnetic instability of the neutral 2SC phase. This mechanism offers a compelling explanation for the origin of the extreme inner magnetic fields of magnetars.

\begin{acknowledgments}
The author wants to express his gratitude to E. J. Ferrer and V. de la Incera for their helpful discussions and insights. This research was funded by the National Natural Science Foundation of China (NSFC), grant number 12075093.   
\end{acknowledgments}

\appendix

\newpage 

\end{document}